%% file: main.tex
\documentclass[twocolumn]{article} 

\usepackage{preprint}

\usepackage{amsmath, amsthm, amssymb, amsfonts}

\usepackage[numbers,square]{natbib}
\usepackage[utf8]{inputenc}	
\usepackage[T1]{fontenc}	
\usepackage[colorlinks = true,
            linkcolor = purple,
            urlcolor  = blue,
            citecolor = cyan,
            anchorcolor = black]{hyperref}	
\usepackage{booktabs} 		
\usepackage{nicefrac}		
\usepackage{microtype}		
\usepackage{lineno}		
\usepackage{float}			

\usepackage{newfloat}
\DeclareFloatingEnvironment[name={Supplementary Figure}]{suppfigure}
\usepackage{sidecap}
\sidecaptionvpos{figure}{c}

\usepackage{titlesec}
\titlespacing\section{0pt}{12pt plus 3pt minus 3pt}{1pt plus 1pt minus 1pt}
\titlespacing\subsection{0pt}{10pt plus 3pt minus 3pt}{1pt plus 1pt minus 1pt}
\titlespacing\subsubsection{0pt}{8pt plus 3pt minus 3pt}{1pt plus 1pt minus 1pt}

\usepackage{tikz,xcolor,hyperref}
\definecolor{lime}{HTML}{A6CE39}
\DeclareRobustCommand{\orcidicon}{
	\begin{tikzpicture}
	\draw[lime, fill=lime] (0,0)
	circle [radius=0.16]
	node[white] {{\fontfamily{qag}\selectfont \tiny ID}};
	\draw[white, fill=white] (-0.0625,0.095)
	circle [radius=0.007];
	\end{tikzpicture}
	\hspace{-2mm}
}
\foreach \x in {A, ..., Z}{\expandafter\xdef\csname orcid\x\endcsname{\noexpand\href{https://orcid.org/\csname orcidauthor\x\endcsname}
			{\noexpand\orcidicon}}
}

\title{In-ear ECG Signal Enhancement with Denoising \\ Convolutional Autoencoders}

\usepackage{authblk}

\makeatletter
\def\@fnsymbol#1{\ensuremath{\ifcase#1\or *\or *\or *\or *\or *\else\@ctrerr\fi}}
\makeatletter

\author[1\thanks{\tt{correspondence}}]{Edoardo Occhipinti\orcidA{}}
\author[2]{Marek Zylinski}
\author[2]{Harry J. Davies}
\author[2]{Amir Nassibi}
\author[2]{Matteo Bermond}
\author[3]{Patrik Bachtiger}
\author[3]{Nicholas S. Peters}
\author[2\thanks{\tt{correspondence}}]{Danilo P. Mandic}

\affil[1]{UKRI Centre for Doctoral Training in AI for Healthcare, Department of Computing, Imperial College London, UK}
\affil[2]{Department of Electrical and Electronic Engineering, Imperial College London, UK}
\affil[3]{National Heart and Lung Institute, Imperial College London, UK}

\begin{document}

\twocolumn[
\maketitle
\vspace{-0.5cm}
\centering
*\{edoardo.occhipinti16, d.mandic\}@imperial.ac.uk \\
\vspace{0.5cm}

\begin{abstract}
The cardiac dipole has been shown to propagate to the ears, now a common site for consumer wearable electronics, enabling the recording of electrocardiogram (ECG) signals. However, in-ear ECG recordings often suffer from significant noise due to their small amplitude and the presence of other physiological signals, such as electroencephalogram (EEG), which complicates the extraction of cardiovascular features. This study addresses this issue by developing a denoising convolutional autoencoder (DCAE) to enhance ECG information from in-ear recordings, producing cleaner ECG outputs. The model is evaluated using a dataset of in-ear ECGs and corresponding clean Lead I ECGs from 45 healthy participants. The results demonstrate a substantial improvement in signal-to-noise ratio (SNR), with a median increase of 5.9 dB. Additionally, the model significantly improved heart rate estimation accuracy, reducing the mean absolute error by almost 70\% and increasing R-peak detection precision to a median value of 90\%. We also trained and validated the model using a synthetic dataset, generated from real ECG signals, including abnormal cardiac morphologies, corrupted by pink noise. The results obtained show effective removal of noise sources with clinically plausible waveform reconstruction ability.
\keywords{Hearables \and In-ear ECG \and Denoising \and Autoencoder}
\end{abstract}

\vspace{0.3cm}
] 


\section{Introduction}
\input{introduction}

\section{Methods}
\input{methods}

\section{Results}
\input{results}

\section{Discussion}
\input{discussion}

\section{Conclusion}
\input{conclusion}

\section*{Acknowledgements}
This work was supported by the USSOCOM MARVELS grant and the UKRI Centre for Doctoral Training in AI for Healthcare grant number EP/S023283/1, in collaboration with the NIHR Imperial Biomedical Research Centre. 

\small
\input{main.bbl}

\end{document}

%% file: introduction.tex
Current wearable technologies available on the market have focused on recording cardiac activity either electrically through ECG or optically through photoplethysmogram (PPG). These technologies have predominantly taken the form of wristbands or chest patches. Wristbands can only provide long-term pulse measurements through PPG, while full-wave ECG characterization is limited as it requires the use of both hands, as seen in devices like the Apple Watch. Chest patches, though better suited for continuous cardiovascular monitoring, are cumberstone for daily use outside the hospital. 

Recent advances into wearables have led to the development of "Hearables", ear-worn devices for vital signal monitoring \cite{goverdovsky2017hearables},\cite{mandic2024multimodal}. Due to their high vascularity and proximity to important signal-emitting organs like the heart, lungs, and brain, as well as the relatively stable position of the head \cite{occhipinti2022artefacts} throughout our daily life, the ears are an ideal choice for continuous remote monitoring. For these reasons, it is estimated that the global Hearables market size is set to reach USD 131.68 billion by 2031 \cite{hearables_report}. 

Hearables have demonstrated their capability in capturing essential physiological signals, including electroencephalogram (EEG) \cite{goverdovsky2016EEG}, ECG \cite{von2017ECG}, and PPG, which have been tested already in several applications including sleep monitoring \cite{nakamura2020sleep}-\cite{hammour2024sleep}, estimation of blood oxygen saturation \cite{davies2020spo2}, blood pressure \cite{Balaji2023bp}, cognitive workload \cite{davies2023cognitive}, glucose level \cite{hammour2023glucose}, and drowsiness \cite{kaveh_drowsiness2024}. 

In particular, in-ear cardiovascular health has been monitored with three main modalities: electrical (ECG), optical (PPG), and audio based technologies. Electrical in-ear ECG signals have been measured either in a cross-head configuration \cite{von2017ECG},\cite{ghena2019ecg} or in a single ear one \cite{yarici2024singleear}. In-ear ECG morphology has been recovered by averaging multiple cardiac cycles, and these average waveforms have been validated against Lead I ECG in terms of the timings of the characteristic waves (P wave, QRS complex and T wave) relative to the R peak and their respective correlation and amplitude ratios \cite{von2017ECG},\cite{yarici2024singleear}. This validation also included a subject previously diagnosed with ventricular bigeminy \cite{ghena2019ecg}. Multi-modal data fusion approaches which combine in-ear ECG with optical ear PPG have also shown great promise in more accurately estimating heart rate \cite{zylinski2024datafusion}, as well monitoring heart rate variability (HRV) to classify mental stress \cite{tian2023HRV}. Alternatively, audio-based solutions using microphones to record the heart sounds from the ear canal have shown great promise in estimating heart rate under motion \cite{butkow2024micHR}. A novel technology known as Audioplethysmography (APG) \cite{google2023apg}, has also been validated for heart rate and HRV estimation. APG involves sending a low-intensity ultrasound probing signal through speakers, which is displaced by the ear canal skin and picked up by a microphone, with the resulting signal closely resembling a PPG waveform.

These technologies have thus proven the feasibility of detecting cardiovascular biomarkers from the ear canal. Nevertheless, for clinical level monitoring and diagnosis, ECG remains the gold standard, and to the best of our knowledge the ability to continuously monitor in-ear ECG with real-time full wave characterization and without averaging remains unattainable. Indeed, while in-ear sensors enhance comfort and wearability compared to traditional chest-mounted electrodes, they suffer from a reduced signal to noise ratio (SNR). This is mainly because the voltage changes due to heart activity are significantly smaller when measured across the head and the ears, unlike when measured at more conventional sites such as the chest or arms \cite{von2017ECG},\cite{yarici2024singleear}. Specifically, the amplitude of the in-ear ECG signal, whether in a cross-head or single-ear configuration, is typically about two orders of magnitude lower than that of chest-recorded ECG signals. Moreover, in-ear ECG recordings are often contaminated with other physiological signals of similar amplitude and overlapping frequency, predominantly EEG, but also electrooculography (EOG), and electromyography (EMG). Interference from artefacts caused by jaw or head movements may also further obscure the cardiac information \cite{occhipinti2022artefacts}.

To best exploit the benefits of a continuous in-ear ECG monitoring solution, there is a need to develop algorithms which recover the ECG information and suppress other non-relevant electrical activity. Recent work by \cite{davies2024deepmatch}, using a deep matched filter model, has demonstrated the capability to identify R-peaks within noisy in-ear ECG signals by training the model using in-ear ECG inputs and Lead I ECG outputs. This study also noted that such a model consistently generates an average cardiac cycle waveform with each detected R-peak. While this method allows for accurate R-peak detection, it is limited in addressing true noise reduction and in the identification of abnormal signal morphologies. 

In our study, we aim to develop a model that effectively denoises in-ear ECG, while accurately capturing the subject-specific morphological characteristics of the ECG waveform. 

\begin{figure}[b]
    \centerline{\includegraphics[width=\columnwidth]{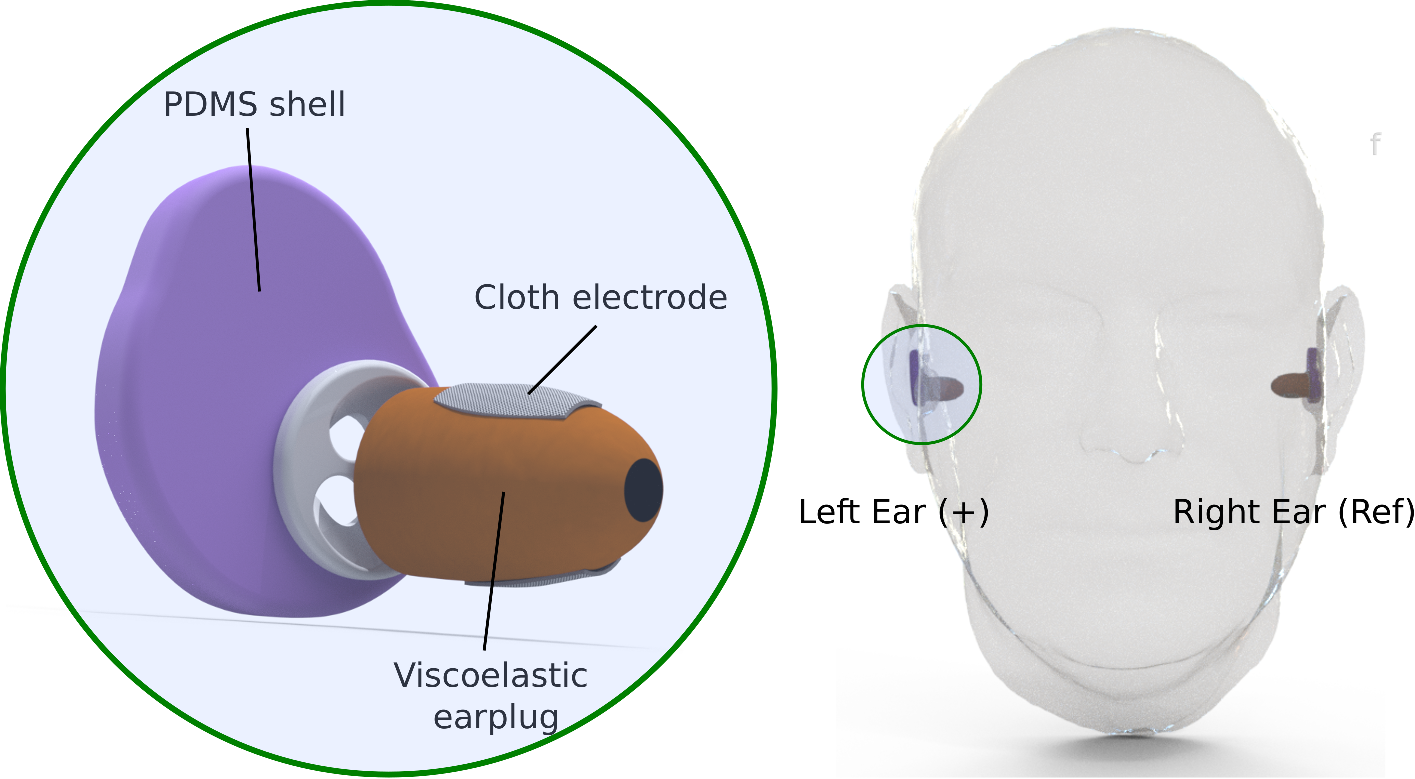}}
    \caption{The earpiece used in this study consists of a gelled cloth electrode attached to a viscoelastic foam and a PDMS shell to anchor it to the concha, thus preventing the earpiece from coming out from the ear canal}.
    \label{fig:earpiece}
\end{figure}

%% file: methods.tex
\subsection{In-ear ECG data collection} \label{sec:ear_ecg_data}
The collected in-ear ECG dataset consisted of simultaneous recordings of cross-head in-ear ECG and Lead I ECG from 45 participants. All subjects were healthy individuals with no know cardiovascular abnormalities. The signals were recorded at a sampling frequency of 500 Hz from two different amplifiers, namely the BrainAmp from BrainVision (North Carolina, USA) and the Somno HD amplifier from Somnomedics (Randersacker, Germany). Each recording lasted for 5 minutes while the subjects were asked to sit and relax. Some artefacts due to eye blinking, head and jaw movements were observed in our recordings, but these data were not excluded from our analysis. For each participant, in-ear ECG measurements were obtained using two earpieces, with one placed in each ear as shown in Figure \ref{fig:earpiece}. Each earpiece, as described in \cite{goverdovsky2017hearables},\cite{mandic2024multimodal}, consisted of a viscoelastic foam earplug with a gelled cloth electrode, held in place by a PDMS shell. Before insertion, the gel was applied to the cloth electrode to ensure optimal contact with the skin and minimize impedance. The research protocol received approval from the IC ethics committee under the reference JRCO20IC6414, and all participants provided written consent prior to starting the recording sessions.

The in-ear ECG recordings were then downsampled at 100 Hz, and segmented in 10 s windows with 90\% overlap. This resulted in a total of 12842 data segments spread across the different subjects.  
A 10s window was chosen as sufficient to encompass multiple cardiac cycles, and such that transient artefacts (e.g. due to blinking) would have less impact on the assessment of denoising effectiveness over the entire signal length. The number of cycles contained in each segment varies based on the individual's heart rate and was deliberately not controlled. The windowed signals, both the in-ear ECG as well as the Lead I ECG, were preprocessed according to the following steps: bandpass filtering 1-40 Hz, 0-1 amplitude scaling, and mean removal. 

\subsection{Noise modelling and data augmentation} \label{sec:eeg_noise}

During training, the dataset was augmented using two distinct methods. The first method involved standard transformations, including inverting the signal and flipping it horizontally, which increased the number of in-ear ECG examples by 25\%. This approach was taken to prevent the model from learning a fixed order of the P, QRS, and T waves, which might not always be present in abnormal ECG signals, where some of these waves may be missing. The inversion was specifically introduced to handle instances where some waves might be inverted, such as inverted T waves. Besides this more conventional transformation, the dataset was also enhanced by incorporating synthetic examples of clean ECG signals artificially corrupted by noise. The clean ECG signals used were extracted from the PTB-XL database \cite{PTBpaper2020},\cite{PTBdata2022}, which includes 21799 recordings, each lasting 10 s, from 18869 participants, with a gender distribution of 52\% male and 48\% female and an age range from 0 to 95 years (median age 62, interquartile range 22). This dataset encompasses both healthy individuals and subjects with arrhythmias, thus further increasing the morphological variability of the ECG signals. For our study, we concentrated specifically on Lead I ECG data because the cardiac vector in this configuration is parallel to that in cross-head in-ear ECG \cite{von2017ECG}, making it most relevant for our research goals. We opted for the 100 Hz version of this dataset to enhance computational efficiency, facilitating future integration into hardware solutions. 

\begin{figure}[t]
    \centerline{\includegraphics[width=\columnwidth]{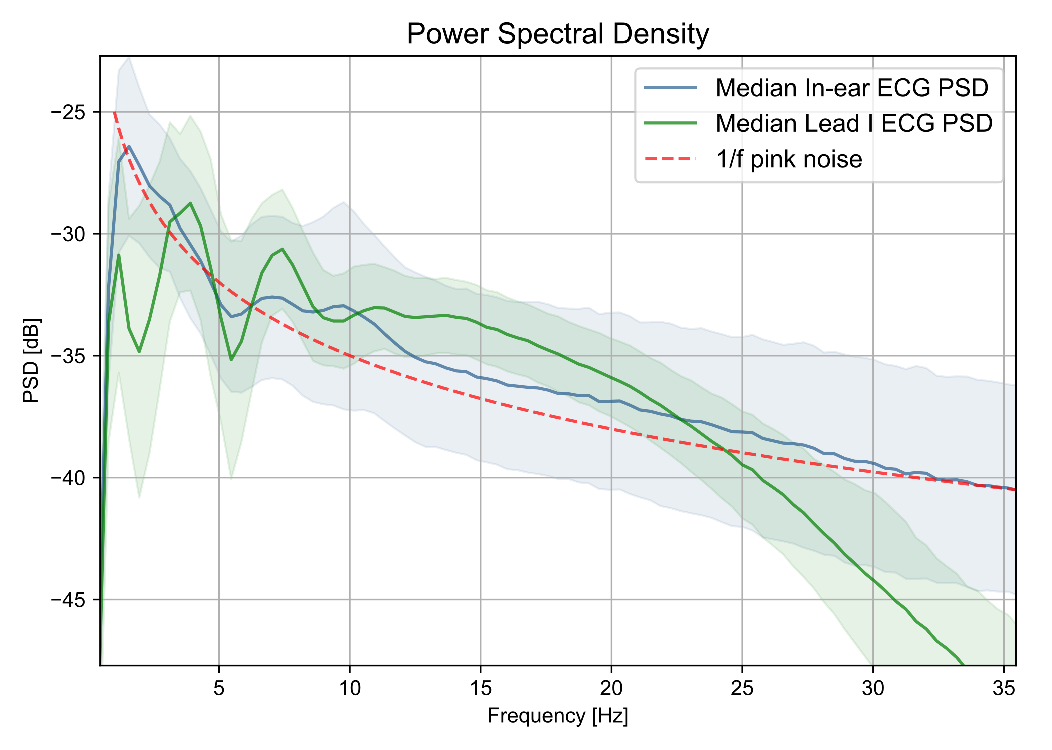}}
    \caption{Average PSD comparison of in-ear ECG and Lead I ECG across the 10 s signals. We can observe how the in-ear ECG spectrum resembles more closely a 1/f spectrum, typical of EEG signals.}
    \label{fig:psd_comparison}
\end{figure}

In order to model the noise, we looked at the power spectral density (PSD) of the in-ear ECG as well as the Lead I ECG (Figure \ref{fig:psd_comparison}). It is evident that, while the the Lead I ECG shows some clear slow-frequency peaks below 7 Hz which correspond to the P and T waves, and a broadband peak up to 30 Hz which corresponds to the QRS complex, the in-ear ECG spectrum is more similar to that of an EEG signal, exhibiting a 1/f frequency distribution. It is also possible to observe a peak at approximately 10 Hz, which corresponds to alpha wave activity, due to most subjects relaxing and closing their eyes during the recordings. We therefore concluded that the main source of noise in in-ear ECG was coming from the EEG signal recorded at the same time. 

\begin{figure}[ht]
    \centerline{\includegraphics[width=\columnwidth]{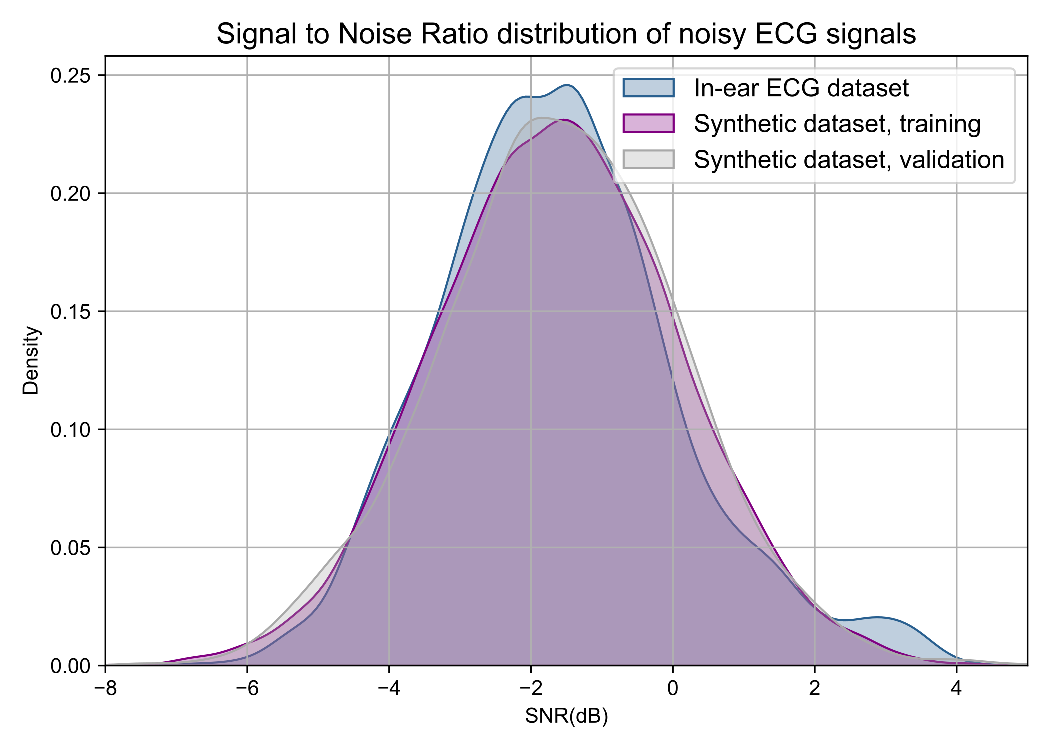}}
    \caption{SNR distribution of in-ear ECG dataset and synthetic dataset generated by corrupting ECG signals from the PTB-XL database with pink noise.}
    \label{fig:SNR_distribution}
\end{figure}

As such, a synthetic dataset was created by combining clean ECG signals with EEG noise. The noise was generated using the MATLAB function \texttt{pinknoise}, sampled at 100 Hz, and bandpass filtered between 1 and 40 Hz to match the preprocessing requirements. This method produces a noise signal with a $1/f$ power spectral density (PSD), characteristic of pink noise and EEG. The noise was added to the preprocessed ECG signal ${\bf{x}}(n)$, such that the SNR (Equation \ref{eq: SNR}) of the noisy ECG signals $\tilde{{\bf{x}}}(n)$ would fall within a similar distribution ($\mu_{SNR} =-1.62$, $\sigma_{SNR} = 1.71 dB$) to the one of the recorded in-ear ECG signals (Figure \ref{fig:SNR_distribution}). This approach ensures that the synthetic noisy examples can be effectively used to train and validate the model under challenging conditions similar to those observed in in-ear ECG recordings.

\begin{equation} \label{eq: SNR}
    SNR = 10 \times log10 \Bigg(\frac{\sum_{n=1}^{N}{\bf{x}}(n)^2}{\sum_{n=1}^{N}({\bf{x}}(n) - \tilde{{\bf{x}}}(n))^2}\Bigg)
\end{equation}

\subsection{1D Denoising convolutional autoencoder}

\begin{figure*}[t]
    \centerline{\includegraphics[width=0.9\textwidth]{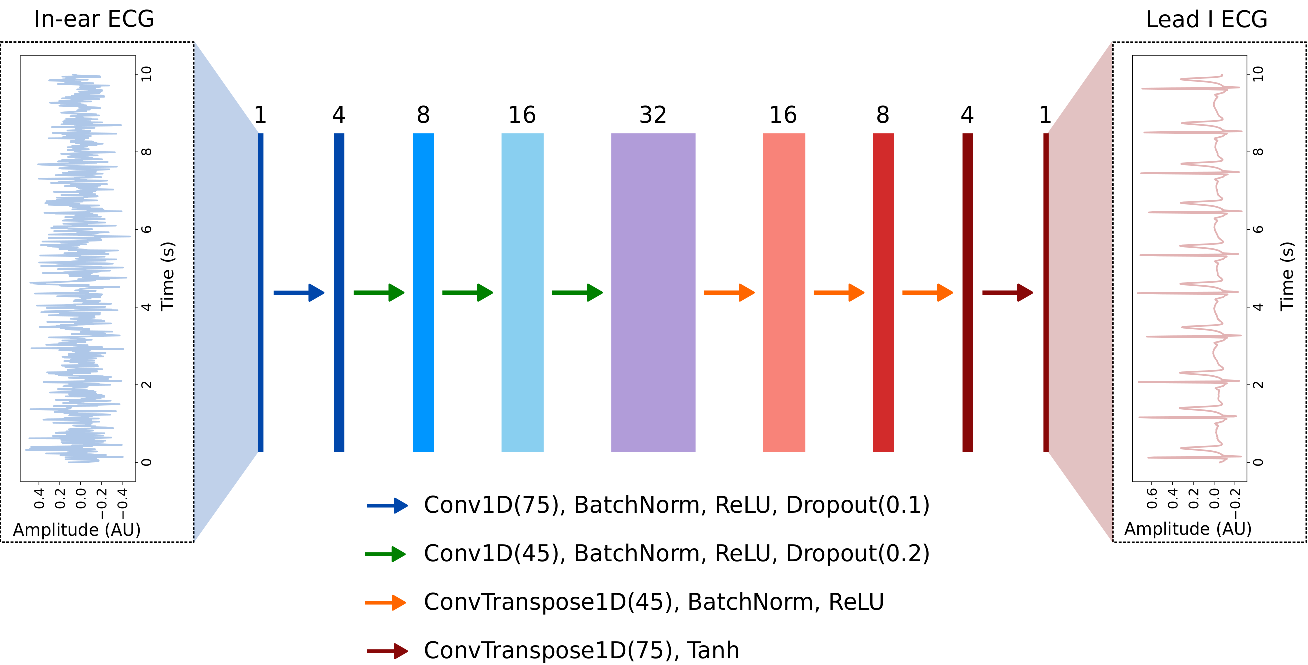}}
    \caption{Denoising convolutional autoencoder (DCAE) architecture taking a noisy in-ear ECG as input, and simultaneously recorded clean ECG as output.}
    \label{fig:architecture}
\end{figure*}

The model architecture to denoise the ECG signals was based on a 1D denoising convolutional autoencoder (DCAE), with the in-ear ECG signals as noisy inputs and the corresponding Lead I ECG as clean outputs. An autoencoder is a type of machine learning model which creates efficient encodings of input data within a latent space representation, and then reconstructs the input from this representation. A natural extension of this is a denoising autoencoder, which aims to reconstruct a clean input from its corrupted version, as shown in Equation \ref{eq:DAE}:

\begin{equation}\label{eq:DAE}
    \hat{{\bf{x}}} = f(g(\bf{x} + \eta))
\end{equation}

where $g$ is the encoder function, $f$ the decoder, and $\eta$ the noise. These models have been explored in various fields, including biosignal processing. For instance, in \cite{denoiseECG2019} a denoising convolutional autoencoder was used to remove white Gaussian noise from Lead I ECG data. 

The proposed architecture, developed in PyTorch \cite{pytorch2019} is shown in Figure \ref{fig:architecture}. The encoder comprises four convolutional layers, with a kernel size of 75 samples (0.75s) in the first layer and 45 samples (0.45s) in the subsequent three layers. Padding was employed in the convolution operations to prevent sample loss, preserving the temporal resolution of the signal. 

The initial layer's kernel size of 0.75 s is sufficient to capture a full cardiac cycle, while the 0.45 s kernel size in the following layers, corresponding to a normal QT duration, is designed to capture dependencies between successive characteristic waves.
The number of channels in each layer was optimised to balance performance and computational efficiency, resulting in a relatively small network with only 61,345 parameters. The convolution operation in each layer was followed by Batch Normalisation, ReLU activation function, and dropout. The decoder was built symmetrically to the encoder, with four transpose convolutional layers. The inner three layers were followed by Batch Normalisation and RelU activation function, while the last one only had a Tanh activation function since the output data is expected to be in the [-1, 1] range after the preprocessing steps. The chosen loss function is the Mean Square Error (MSE) which is adequate for a denoising task. 

\subsection{Model evaluation}
The pairs of noisy/clean ECG across the two datasets (in-ear ECG and PTB-XL) were then mixed and divided into training, validation and test datasets. The training dataset comprised a mix of synthetic examples (folds 1-8 in the PTB-XL database), and in-ear ECG examples, while the validation dataset contained only synthetic examples (folds 9-10 in the PTB-XL database). To rigorously evaluate the denoising performance and ensure model robustness and generalizability, a leave-1-subject-out testing approach was used. In fact, a separate test dataset was built for each model using in-ear ECG data from one previously unseen subject, and all results were estimated from evaluating the model on this subject.

Each model was trained with a batch size of 32 for a maximum of 20 epochs using early stopping to prevent overfitting (i.e., stopping when the testing loss ceased to improve).

%% file: results.tex
\subsection{Model Behaviour}

\begin{figure}[hbt]
    \centerline{\includegraphics[width=\columnwidth]{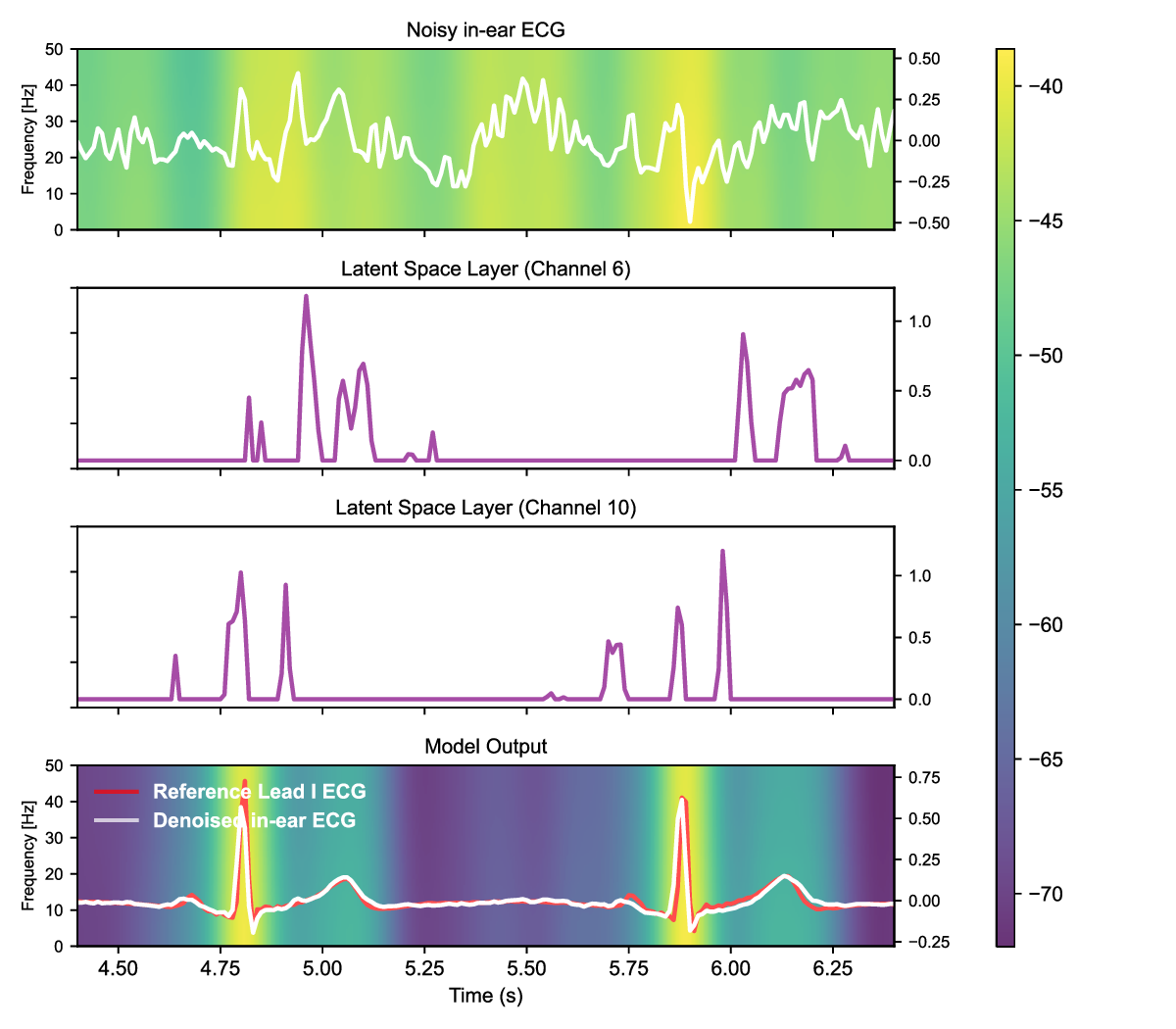}}
    \caption{Top panel: Zoom-in view of the input in-ear ECG showing two successive cardiac cycles, with the corresponding spectrogram in the background. Middle panels: Latent space representation of 2 out of 32 channels, illustrating how different morphological features of the ECG are captured. Bottom panel: Denoised in-ear ECG along with the associated spectrogram, displayed alongside the reference Lead I signal. The spectrogram was computed using a 0.1s window with 50\% overlap and smoothed using a Gaussian filter.}
    \label{fig:spectrogram}
\end{figure}

To illustrate the working principle of the model, Figure \ref{fig:spectrogram} shows the process of compressing a typical in-ear ECG waveform (top panel) into the Latent Space and then expanding it back into a clean, denoised ECG waveform (bottom panel).

The middle panels display two channels from the Latent Space, revealing that they each focus on different regions of the ECG template. Channel 6 captures more T-wave information, while Channel 10 concentrates on the P-wave and QRS complex. This analysis confirms that the model is indeed examining various morphological aspects of the ECG, rather than merely focusing on the R peak (which could disproportionately influence the MSE loss) and learning a fixed wave pattern.

The spectrograms plotted as backgrounds in the top and bottom panels also help us understand how the denoising is successfully eliminating the spectral components which are not associated with ECG.

\subsection{Denoising performance evaluation}
To evaluate a successful denoising, we looked at how the SNR changed from the original input signal to the denoised version. We refer to this as a SNR improvement: 
\begin{equation}\label{eq:SNR_imp}
    SNR_{imp} = SNR_{out} - SNR_{in}
\end{equation}

\begin{figure}[hbt]
    \centerline{\includegraphics[width=\columnwidth]{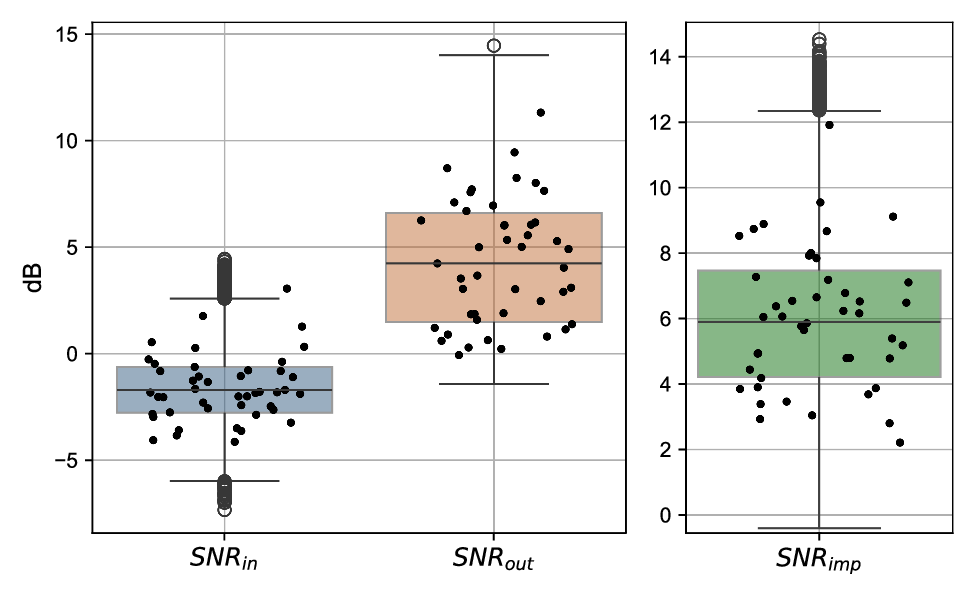}}
    \caption{(a) SNR of input in-ear ECG signals ($SNR_{in}$) and SNR after denoising ($SNR_{out}$); (b) SNR improvement ($SNR_{imp}$). Each dot represents the average SNR for each subject.}
    \label{fig:snr_imp}
\end{figure}

We evaluated the different models on each test dataset and reported the results for the unseen subjects. The dots in the plot represent the average SNR for each subject. We observed that only 1 out of the 45 subjects has a negative average $SNR_{out}$ after denoising compared to $SNR_{in}$ where 39 subjects initially had a negative SNR. The SNR improvement yielded a median value of 5.9 dB and consistently demonstrated a positive increase of at least 2dB on average per subject, as shown in Figure \ref{fig:snr_imp}b.

\subsection{Heart Rate estimation in-ear ECG}
To further evaluate the benefits of denoising, we examined how heart rate estimation from in-ear ECG signals is affected after denoising. The Matched Filter Hilbert Transform (MFHT) algorithm \cite{chan2015mfht} was used to compare heart rate estimation performance between the original preprocessed in-ear ECG signals and the denoised ones. This method was specifically developed to estimate heart rate under noisy conditions, like the ones observed in in-ear ECG measurements. The MFHT algorithm calculates the envelope $a$ of the Hilbert Transform of the cross-correlation $y(n)$ between the ECG signal $x(n)$ and an ECG template $w(k)$ extracted by averaging across the cardiac cycles of 1 of the test subjects. 

\begin{equation}\label{eq:cross_correlation}
    y(n) = \sum_{k}{x(n+k)w(k)} 
\end{equation}

If we consider the cross-correlation $y(n)$ in Equation \ref{eq:cross_correlation}, and we denote $y^h(n)$ as the Hilbert Tranform, the envelope $a(n)$ can be extracted according to:

\begin{equation}\label{eq:envelope}
a(n) = \sqrt{(y^2(n) + y^{(h)2}(n)}
\end{equation}

\begin{figure}[t]
    \centerline{\includegraphics[width=\columnwidth]{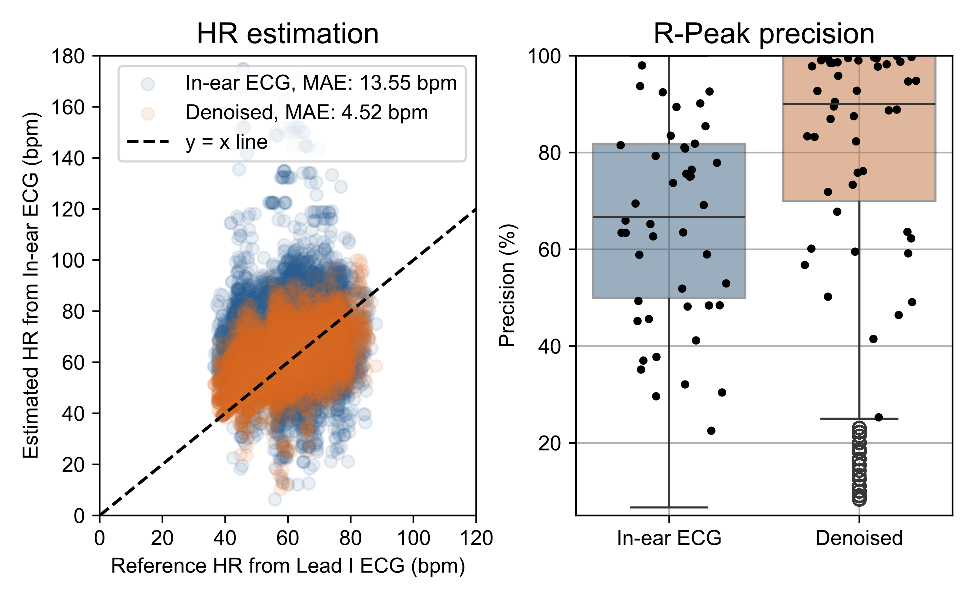}}
    \caption{Left: Scatter plot of estimated Heart Rates from in-ear ECG for all test signals. The MAE is reduced by 67\% after denoising; Right: Precision of R-peak detection on in-ear ECG signals before and after denoising. Each dot represents the average precision for each subject.}
    \label{fig:HR_estimation}
\end{figure}

The Heart Rate was then estimated using the \texttt{scipy.signal.findpeaks} function to identify the peak location from $a$. The minimum distance was set to 333ms (sufficient to capture up to 180 bpm), and a minimum height of twice the mean of the envelope were set as parameters. The results are shown in Figure \ref{fig:HR_estimation}. 

The Heart Rate estimation was evaluated both in terms of the Mean Absolute Error (MAE) between the in-ear ECG and the reference Lead I ECG, and in terms of the precision of the R-peak location with a tolerance of 60ms (half the length of a normal QRS complex). The MAE was reduced from 13.55 bpm to 4.52 bpm which corresponds to a 67\% error reduction. Furthermore, the estimated heart rates from the denoised in-ear ECG are more clustered around the diagonal $y=x$ line, indicating an improved correlation. 

Regarding the R-peak precision, the median value increased from 66.6\% to 90.0\%, and in all subjects the average R-peak precision (dots in Figure \ref{fig:HR_estimation}-right) has improved significantly with a p-value of $P = 6\times10^{-5}$ from a one-way ANOVA test. The R-peak detection performance is comparable to that of the previously developed deep-matched filter model \cite{davies2024deepmatch}.

\begin{figure*}[t]
    \centerline{\includegraphics[width=\textwidth]{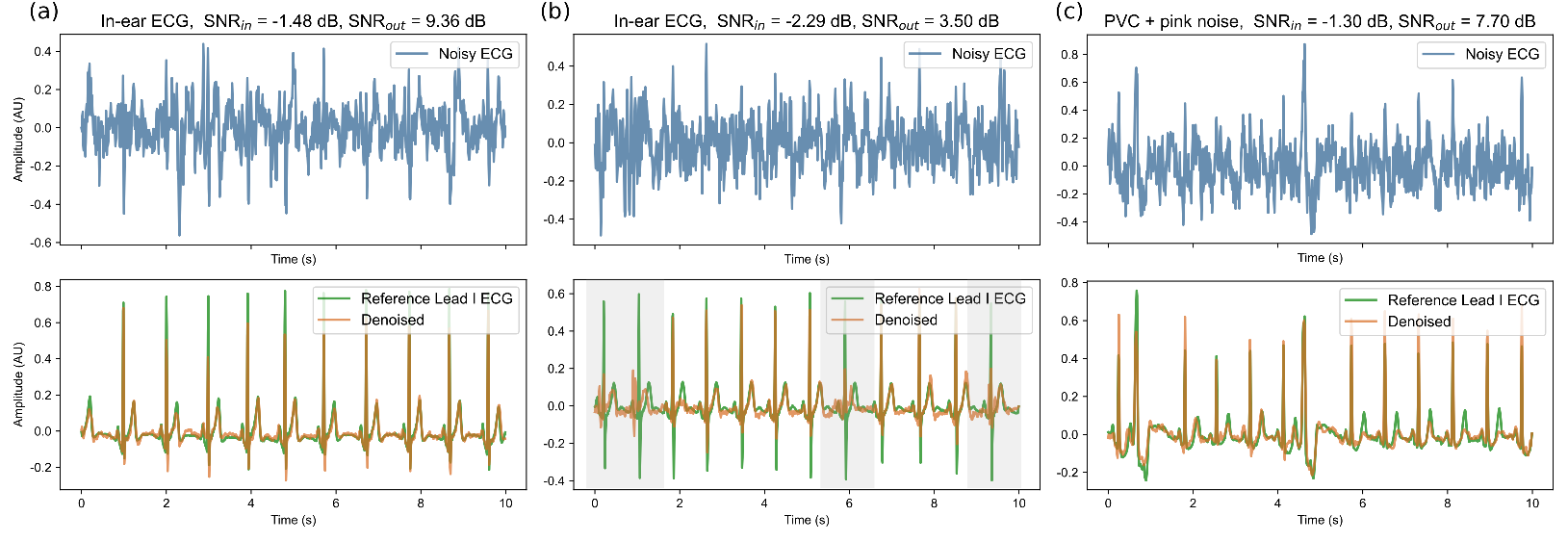}}
    \caption{(a)-(b) in-ear ECG denoising examples from different subjects with different input SNR. The highlighted areas shows those cases where the morphology could not be fully recovered; (c) Validation example on a Premature Ventricular Contraction (PVC) ECG corrupted with pink noise to prove the ability of the model to capture the real underlying abnormal ECG morphology.}
    \label{fig:denoising examples}
\end{figure*}

%% file: discussion.tex
The results in this study show that while recording electrical signals in a cross-head configuration (Figure \ref{fig:earpiece}), a mix of different physiological signals is captured, including in-ear ECG \cite{von2017ECG} and in-ear EEG \cite{goverdovsky2016EEG}. For the purpose of extracting cardiovascular biomarkers from in-ear recorded signals, EEG represents a source of noise which should be removed in order to increase the SNR and to reconstruct clinically plausible ECG waveform signals. The implementation of a DCAE trained with noisy in-ear ECG signals and synthetic ECG+EEG signals as inputs, and the clean Lead I ECG signals as outputs, has significantly improved the SNR across different individuals with varying initial signal quality. The model's ability to maintain the morphological integrity of ECG, even with noise, is critical for any possible application of hearables in clinical practice. 

Figure \ref{fig:denoising examples} illustrates the model's performance on both in-ear ECG recordings from healthy participants, as well as more complex morphologies artificially corrupted by pink noise. In the input noisy signals (top panel), the high contamination of signal with noise makes the extraction of cardiovascular information challenging. After denoising, the model shows a high degree of fidelity in capturing essential features, by correctly reproducing an ECG signal with a higher SNR, making it much easier to be interpreted by cardiologists. The performance of denoising depends on the input signals quality, with some cases almost perfectly matching the reference Lead I ECG (Figure \ref{fig:denoising examples}a), and others struggling to reconstruct these features (highlighted areas in Figure \ref{fig:denoising examples}b) in some cardiac cycles. This difference may be observed in the presence of artefacts occurring during recordings, at an intra-subject level, or due to different physiological characteristics such as heart orientation or other conditions, at an inter-subject level. Indeed, the propagation of the heart’s cardiac dipole when recorded at the ear level has not been fully examined relative to different heart orientations and even different body mass indices (BMI). These are well-known factors in electrophysiology studies, and they are one of the key reasons why ECG is usually monitored with a 12-lead configuration on the chest, which allows capturing the heart’s cardiac cycle from different angles.

We have also tested the model on our synthetic validation dataset on cases where certain more complex morphologies are present, such as those observed in premature ventricular contraction (PVC), shown in Figure \ref{fig:denoising examples}c. Even in this scenario, the model is able to successfully capture the abnormal cycles with high fidelity, which is promising for its application in real-world clinical scenarios where such abnormalities may be present.

We also aimed to ensure that the model could robustly capture the different morphological components of the ECG signal independently. Specifically, we wanted to verify that the presence of a QRS complex does not always imply a subsequent T wave, or vice versa. This independence is important in cases of specific arrhythmias, where the T wave might be flattened or merged with a wider QRS complex. To test the model's robustness in such scenarios, we occluded certain regions of the ECG signal with White Gaussian Noise (WGN) of similar amplitude to the rest of the input signal.

\begin{figure}[hbt]
    \centerline{\includegraphics[width=\columnwidth]{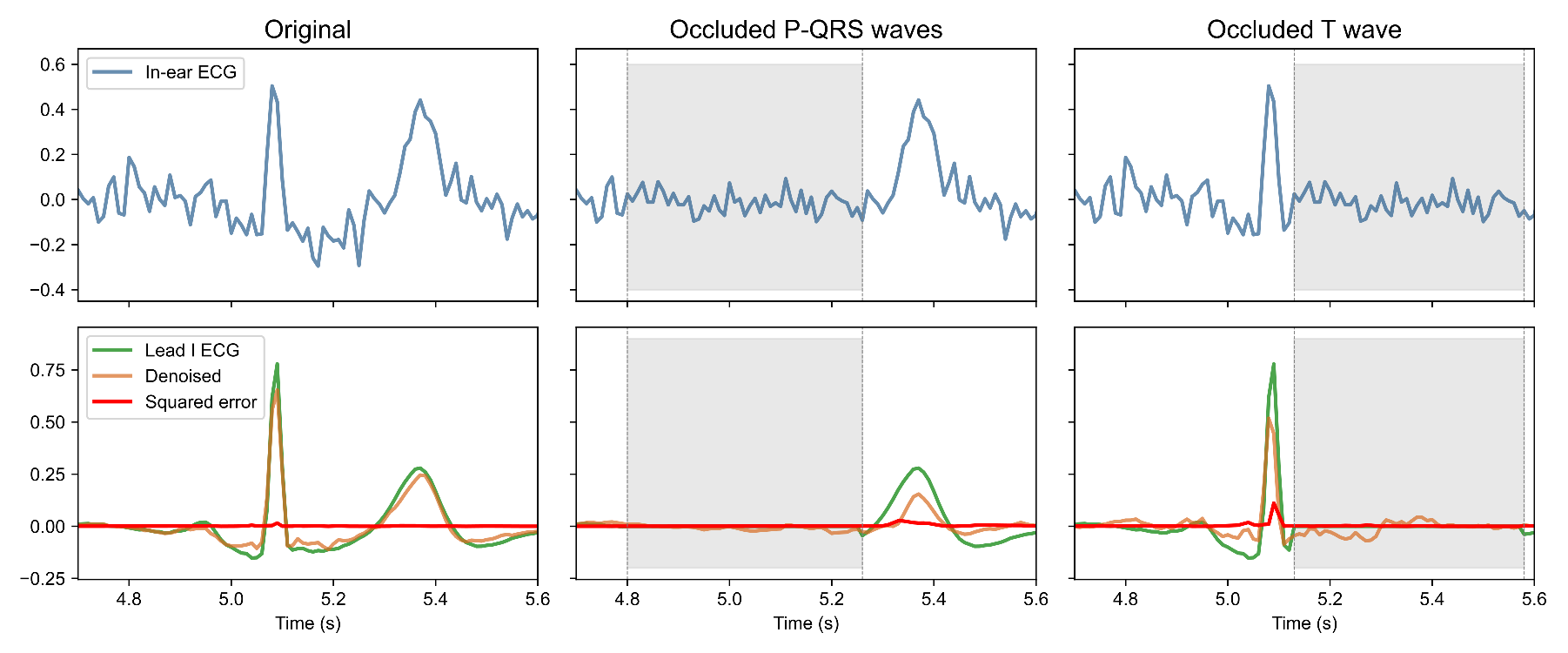}}
    \caption{Left panel: Original in-ear ECG cardiac cycle denoising; Middle panel: Occlusion of the P-QRS portion of the cycle with WGN; Right panel: Occlusion of the T-wave with WGN.}
    \label{fig:occlusion}
\end{figure}

Figure \ref{fig:occlusion} illustrates this test. In the middle panel, we occluded the initial part of the ECG (P-wave and QRS complex), and in the right panel, we occluded the final part (T-wave). The denoised version of the signal did not automatically reproduce a full cardiac cycle when only one of these features was present. Additionally, we observed that the model predicted a smaller amplitude for both the T wave and QRS complex, reflecting its reduced confidence that the input signal was indeed an ECG.

These examples highlight the model's robustness in enhancing ECG signal quality, which is crucial for detecting both typical (healthy) and atypical (potentially pathological) cardiac events.

%% file: conclusion.tex
The findings of this study successfully demonstrate that denoising convolutional autoencoders (DCAE) can significantly enhance in-ear ECG signals, making them comparable to conventional Lead I ECGs. The implementation of DCAE significantly improved the SNR, heart rate estimation accuracy, and R-peak detection precision across different subjects, highlighting its robustness and potential for clinical applications. The model's ability to maintain the morphological integrity of ECG signals, even in the presence of noise, is critical for the potential use of hearables in continuous cardiovascular monitoring, towards towards better detection and earlier intervention for heart disease.

The advancements presented in this study position in-ear ECG as a viable and effective tool for cardiovascular health monitoring, with further implications to be investigated through clinical studies.